\RequirePackage{fix-cm}
\documentclass[twocolumn]{svjour3}          
\smartqed  
\usepackage{mathptmx}      

\usepackage[english]{babel}
\usepackage[applemac]{inputenc}

\usepackage{graphicx}
\usepackage{amsmath}
\usepackage{lipsum}
\usepackage{color}
\usepackage{siunitx}
\usepackage{authblk}
\usepackage{subfigure}
\usepackage[margin=2cm]{geometry}

 \newcommand{\AS}[1]{{\color{black} #1}}

\begin{document}

\title{Influence of the size of the intruder on the reorganization of a 2D granular medium}



\author{Aymeric Merceron $^1$         \and
        Alban Sauret $^{1,2}$ \and
				Pierre Jop $^{1,*}$
}

\authorrunning{Aymeric Merceron, Alban Sauret, Pierre Jop} 

\institute{$^1$
              Surface du Verre et Interfaces, UMR 125, CNRS/Saint-Gobain, 93303 Aubervilliers, France \\
              $^2$
              Department of Mechanical Engineering, University of California, Santa Barbara, USA\\
					\emph{corresponding author:} \email{pierre.jop@saint-gobain.com}  
}

\date{Received: date / Accepted: date}

\maketitle

\begin{abstract}
We consider the rearrangements of a vertical two-dimensional granular packing induced by the withdrawal of an intruder. Here, we focus on the influence of the size of the intruder on the reorganization process. The long term evolution of the granular packing is investigated as well as the avalanche dynamics that characterize the short term rearrangements around the intruder. For small enough intruder, we observe the formation of arches that periodically destabilize and influence the reorganization dynamics of the two-dimensional packing through larger rearrangement events.  \\
\end{abstract}

\section{Introduction}


The description of the evolution of a granular medium submitted to a solicitation often relies on a statistical characterization, for instance through the spatial and temporal evolution of a localized perturbation. The characteristic length over which the perturbation propagates is generally related to the amplitude of the perturbation and the contact force network. Different methods of perturbations of a granular media have been considered. For instance, the translating motion of an intruder can be used to relate the mean displacement field to the local rheological properties \cite{AM105}. However, the mean behavior does not account for the unsteadiness of the response of the granular medium in the vicinity of the intruder. Indeed, the motion of the intruder leads to local reorganizations of the granular packing: the breaking and local reformation of the force network around it \cite{geng2005slow,seguin2016clustering,kolb2013rigid}. The intruder is thus subject to large force fluctuations \cite{AM110}. If the size of the intruder increases, the intruder is more likely to make contact with grains that participate in the strong network, therefore, leading to a larger mean drag force. The geometry of the intruder also has a crucial role. Varying the size of the intruder is a good way to investigate the jamming transition in a granular media \cite{AM112}. The situation becomes even more complicated when considering the motion of many intruders that can then present attractive or repulsive effects \cite{AM114}.

A related situation is the flow of grains in a silo that involves the reorganization of force networks. The granular medium flows under its weight and is guided in an outlet. Providing that the orifice is large enough, the flow reaches a stationary state, and the velocity fields can be modeled \cite{AM115,AM116}. However, when reducing the size of the orifice to a few grains, the formation of arches at the outlet leads to clogging events \cite{AM119}. The larger the orifice is, the smaller the probability of clogging is. It is common to define an avalanche size that corresponds to the number of grains exiting the silo between two clogging events. For the range of orifice sizes that allows intermittent regime, this parameter shows a distribution that decreases exponentially and was attributed to a constant probability of clogging \cite{AM123}. The mean avalanche size shows a divergence beyond a critical width of the orifice \cite{AM121,AM124}. More recent studies exhibited an exponential trend of the avalanche size with respect to the orifice width \cite{AM125,AM126}.

 All of these examples illustrate the need to characterize the spatial and temporal reorganization in a granular assembly submitted to a perturbation. Here, we investigate the quasi-static downward motion of an intruder in a 2D granular media. \AS{In a previous study, the rearrangements of a 2D granular medium undergoing a localized perturbation induced by the withdrawal of a large intruder has been considered \cite{merceron2016reorganization}. With intruder of width much larger than the size of the grains, the long-term displacement is similar to those from 2D quasistatic silos. In addition, the study of the short term rearrangements has shown that the size distribution of the largest events follows a power law. Such behavior is characteristic of processes ruled by scale invariance. However, the influence of the intruder size on the dynamics of rearrangement and the exponent of the power law remained to be explored. For this reason, we focus here} on the influence of the size of the intruder on the rearrangement dynamics of the grains. In particular, the presence of arches can significantly modify the size and the distribution of the reorganization events. We first describe in section \ref{Sec2} the methods. Section \ref{Sec3} is devoted to the longtime rearrangement in the granular packing. We then characterize the short time rearrangement in section  \ref{Sec4} and highlight the influence of the size of the intruder on the reorganization events. Depending on the size of the intruder, the formation of arches can be observed in the system and are reported in section  \ref{Sec5}. 


\section{Methods} \label{Sec2} 

\begin{figure}
\centering
  \subfigure[]{\includegraphics[width=0.5\textwidth]{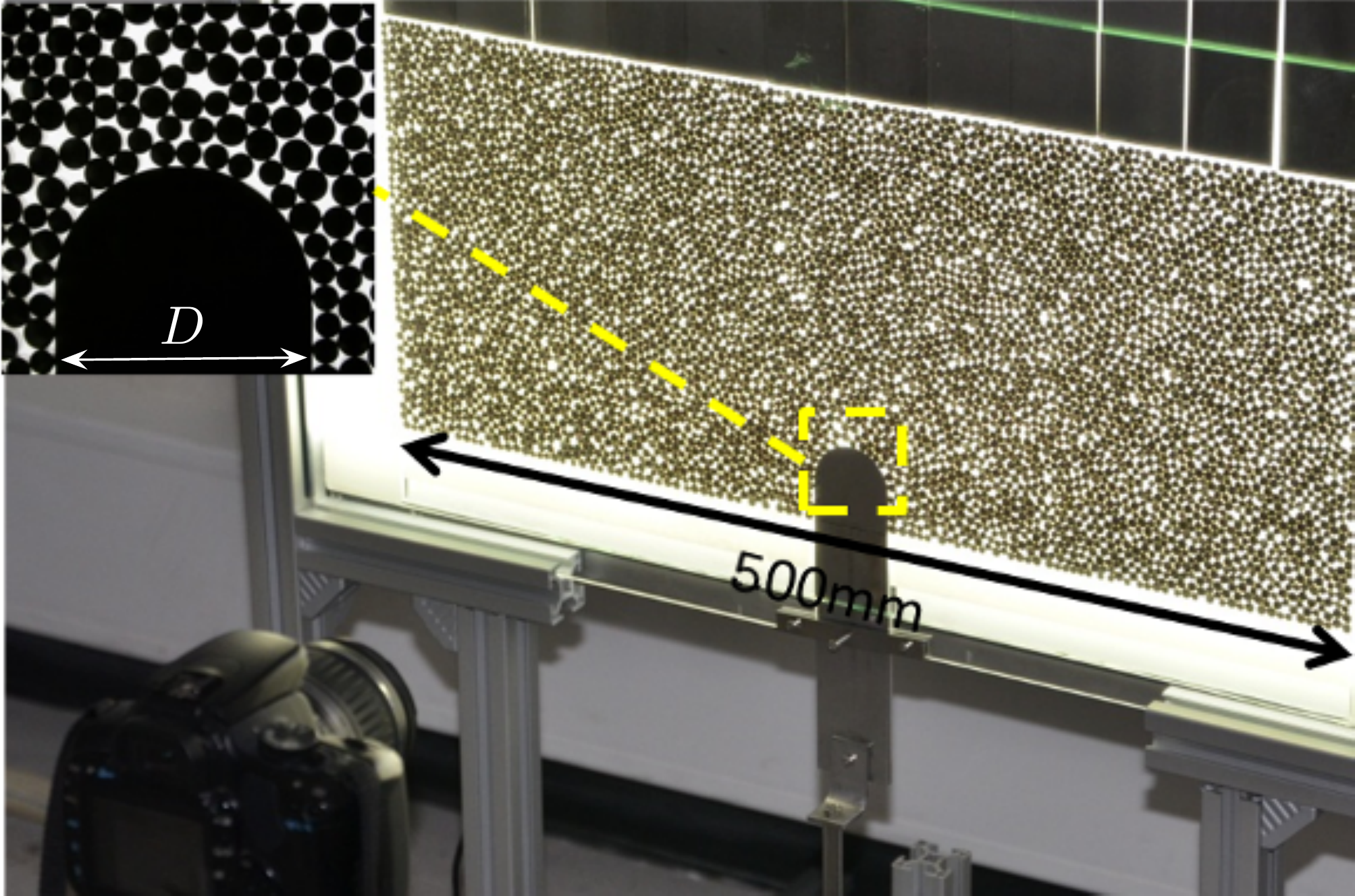}}
    \subfigure[]{\includegraphics[width=0.35\textwidth]{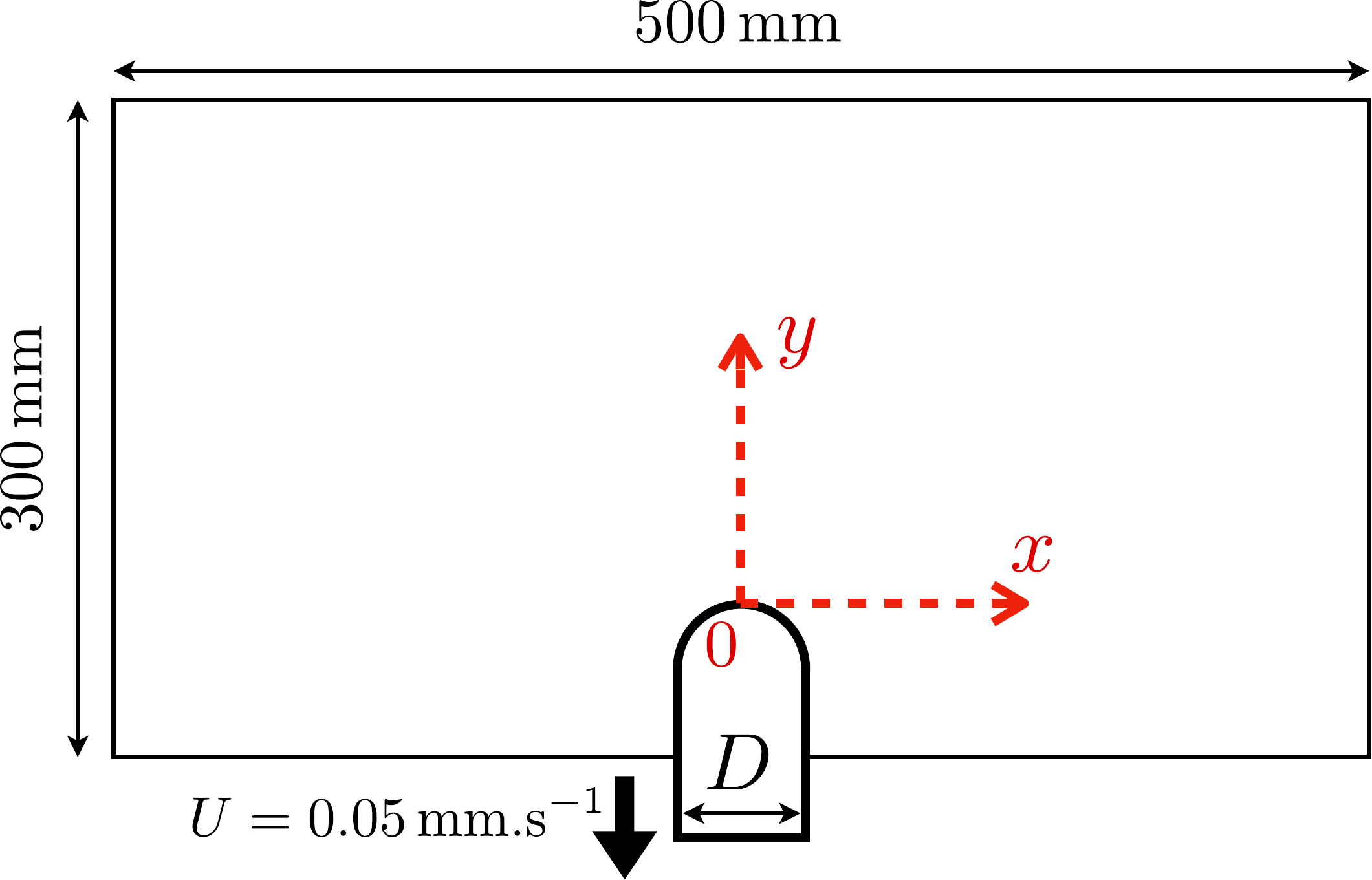}}
 \caption{(a) Photo and (b) schematic of the experimental setup showing the 2D granular media and the mechanical intruder of width $D$.}\label{figure1_setup}
\end{figure}

The experimental setup, shown in Fig. \ref{figure1_setup}(a)-(b), is similar to the method used by Merceron \textit{et al.} \cite{merceron2016reorganization,merceron2018cooperative}. \AS{In summary,} the reorganization of the 2D granular packing is induced by slowly withdrawing a mechanical intruder of width $D$ vertically \AS{using a stepper motor}. The 2D granular packing consists of approximatively 5500 disks of 4 and 5 mm disks (proportions 10:7) placed in a cell made of two parallel glass plates (300 mm high and 500 mm wide). \AS{The disks (nickel-plated brass) are mixed and horizontally inserted in the cell at the same moment as the intruder to form a dense disordered packing. Prior to an experiment, the cell is tilted to a vertical position}. The vertical downward motion induces the reorganization of the granular medium at a low and constant speed of $0.05\,{\rm mm. \,s^{-1}}$ of the mechanical intruder. The intruder has a quasi-circular end (Fig. \ref{figure1_setup}) and a width in the range [10, 40] mm. In all the situations, the downward motion of the intruder is equal to $40\,{\rm mm}$. The slow withdrawing velocity ensures a quasistatic evolution of the granular assembly. The packing is constrained at the top by confining weights ensuring a uniform pressure over the top of the granular assembly. In the following, all lengths are made dimensionless using the diameter of the smallest grains, $d_g = 4\,{\rm mm}$. \AS{
Pictures of the 2D granular packing are recorded at a time interval of $\delta t=5\,{\rm s}$ using a high-resolution camera (Nikon D7000). Thus, the displacement of the
intruder between two pictures is $0.25 mm = d_g/16$.
The pictures are then processed with a segmentation protocol leading to an accuracy smaller than $33\,\mu{\rm m}$) on the position of the disks. Tracking techniques are then used to extract the short-term and long-term
displacements}. To obtain reliable statistical data, we repeat each experiment approximatively 40 times for all configurations.

We also perform numerical simulation using LMGC90, a discrete element software to explore more configurations \cite{dubois2011lmgc90,merceron2016reorganization,merceron2018cooperative}. We measured experimentally the particle-parti-cle and particle-wall dynamical friction coefficients, $\mu=0.13$, and used the same value in the numerical simulation. The mechanical intruder evolves by successive discontinuous steps ($\delta y$ = -1/16 $d_g$ per step) ending when the entire system reaches an equilibrium configuration. The computational time step is set to $10^{-4}\,{\rm s}$, such that the overlap between disks is negligible. The simulations were repeated between 20 and 50 times for each configuration to obtain reliable statistics.

\section{Long term evolution}  \label{Sec3}

\begin{figure}
\centering
	\includegraphics[width=0.48\textwidth]{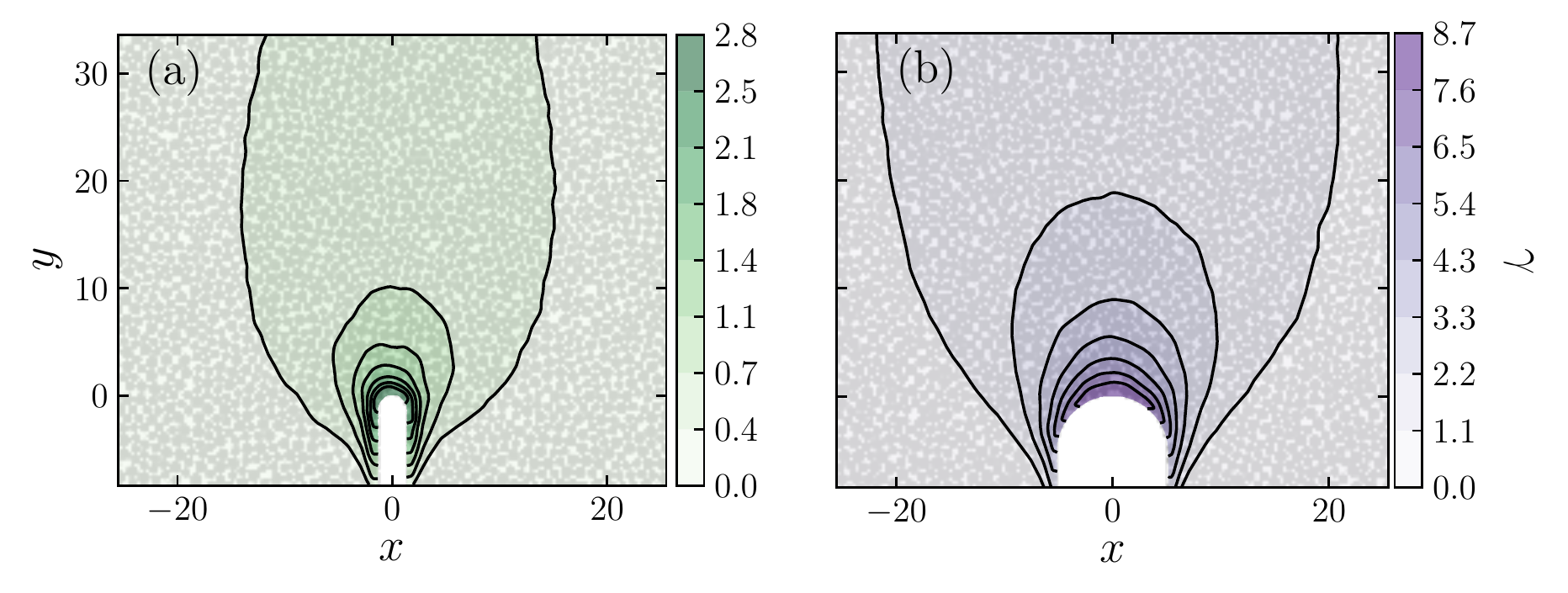}
	\includegraphics[width=0.44\textwidth]{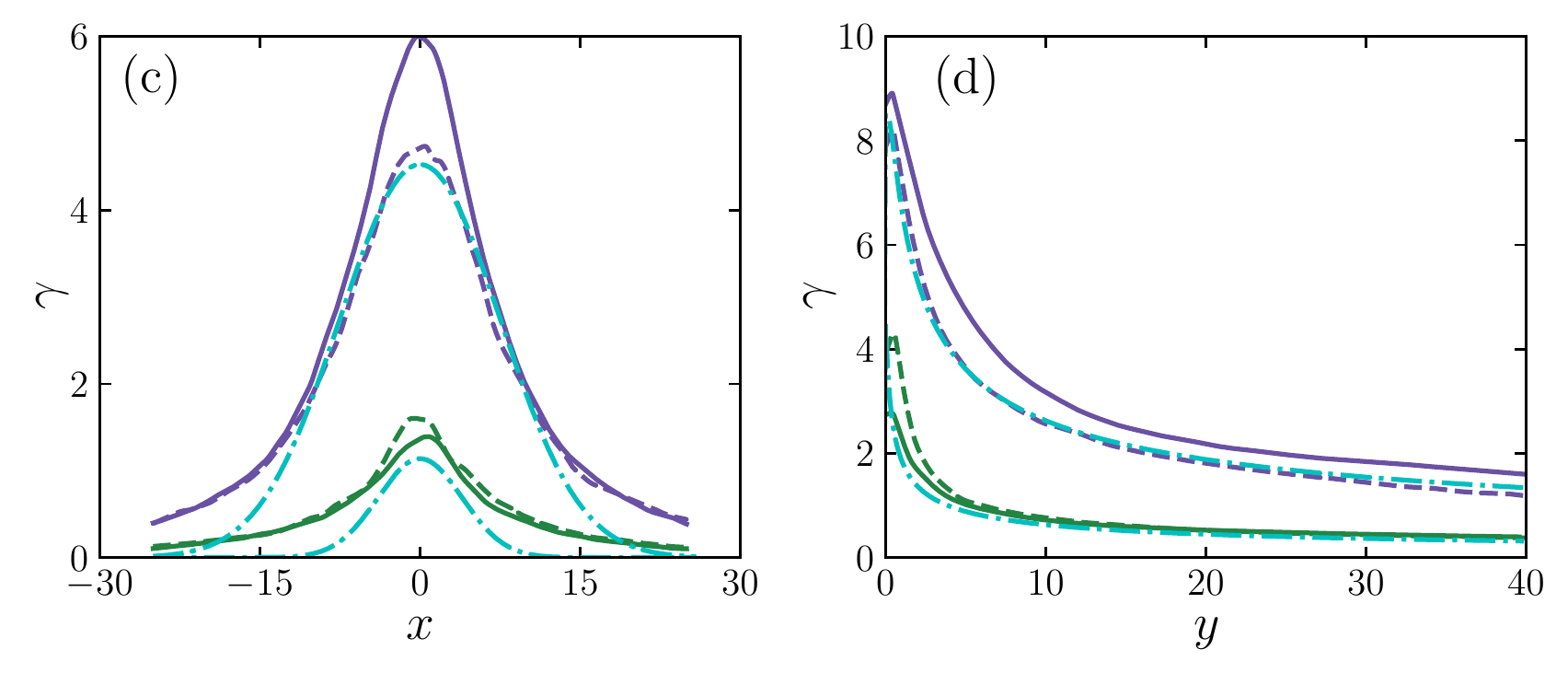}
 \caption{Vertical displacement fields between the initial and final state, \AS{as a function of the initial position}, averaged over 100 experiments with an intruder of width (a) $D=2.5\,d_g$ and (b) $D=10\,d_g$. Plot of the vertical displacement fields $\gamma$ between the initial and final state for the experiments (continuous line) and the numerical simulation (dotted line) along the axis (c) $y = 3$ and (d) $x = 0$. . Purple and green curves are for intruders of width  $D=10\,d_g$ and $D=2.5\,d_g$, respectively. \AS{The light-blue-mixed-lines are the theoretical displacements given by the fit of the Eq.~\ref{eq_1}}}\label{Fig2_LongTime}
\end{figure}

We first investigate experimentally and numerically the displacements of the disks between their initial and final position. The average vertical displacement fields, \AS{denoted $\gamma$, are} obtained experimentally for intruders of width $D=2.5\,{\rm d_g}$ and $D=10\,{\rm d_g}$ and are shown in Fig. \ref{Fig2_LongTime}(a)-(b). The profiles are qualitatively similar and exhibit a strong vertical orientation induced by the gravity. The lateral extent of the region where particles present a strong displacement is related to the intruder width. Besides, although the running length of the intruder is the same, the amplitude of displacement increases with the intruder width. For a small intruder, compared to the size of the disks ($D=2.5\,{d_g}$), stable structures appear, referred to as arches, made by the nearest disks to the intruder. As the arch collapses at its basis, the disk initially located at the top of the arch remains trapped further away from the intruder tip. The numerical simulations performed for the same intruder widths exhibit displacement amplitudes that are comparable as can be seen in Fig. \ref{Fig2_LongTime}(c)-(d).

The long-time behavior of the different configurations presented here can be compared to quasi-static flows in vertical silos. The motion of the granular assembly can be described through a diffusive kinematic model \cite{nedderman_kinematic_1979,melo_kinematic_2008}. This approach describes the proportionality between the horizontal displacement fields and the gradient of the vertical displacement fields along the horizontal axis using a diffusion coefficient $D_p$. This assumption is based on the fact that the disks are more likely to move laterally towards the fastest vertical flow zones. Assuming that the variations of density are sufficiently small to consider the granular medium as incompressible, it is then possible to obtain an analytical expression of the vertical displacement field $\gamma(x,y)$ as a function of the maximum vertical displacement $\gamma_0$ at the orifice (here at the mechanical intruder):
\begin{equation} \label{eq_1}
\gamma(x,y)=-\frac{\gamma_0}{2}\,\left[\text{erf}\left(\frac{x+D/2}{2\,\sqrt{D_p\,y}}\right)-\text{erf}\left(\frac{x-D/2}{2\,\sqrt{D_p\,y}}\right)\right],
\end{equation}
where $x$ and $y$ are the horizontal and vertical coordinates, respectively \AS{(see Fig. \ref{figure1_setup}b)}, \AS{$D/2$ is the half-width of the intruder and $D_p$ is the diffusion coefficient that relates the horizontal and vertical components of the displacement.}

\begin{figure}
\centering
  \includegraphics[width=0.48\textwidth]{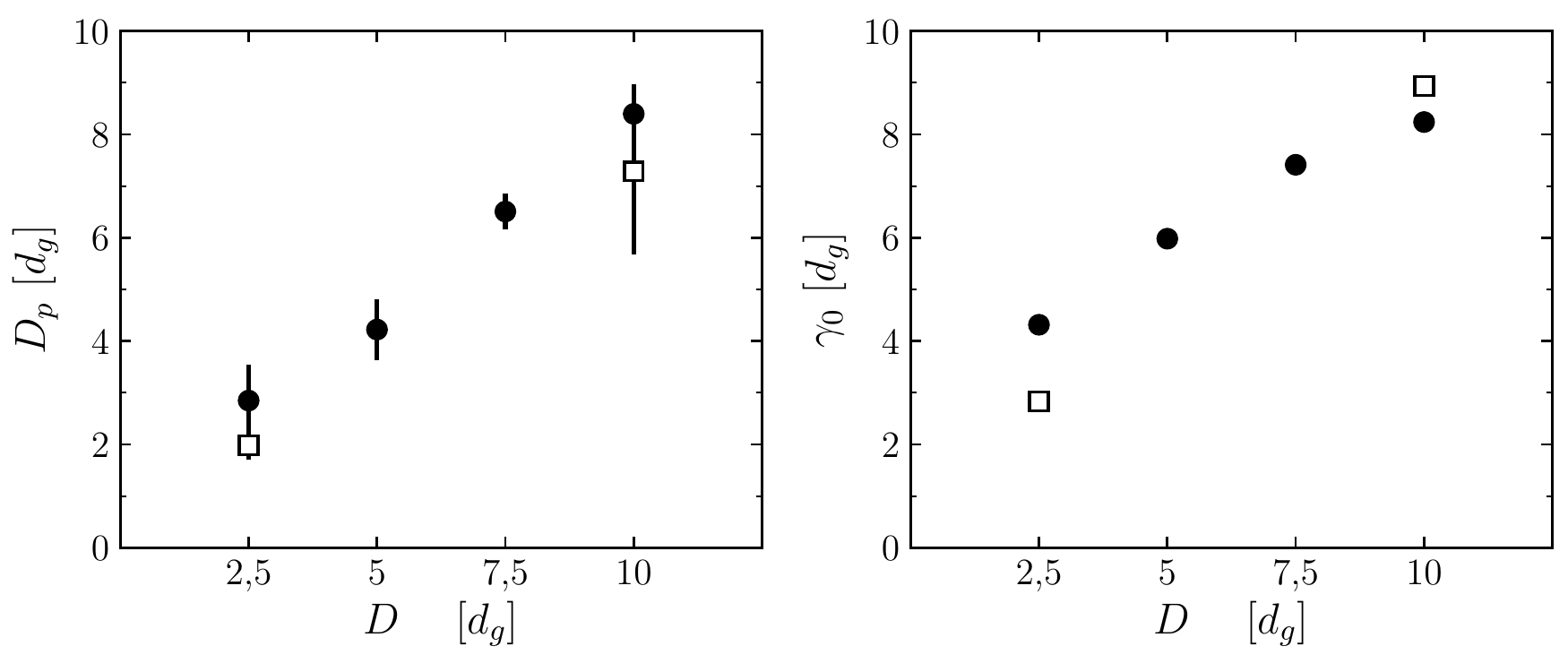}
 \caption{(a) Diffusion coefficient $D_p$ and amplitude of the maximum vertical displacement $\gamma_0$ obtained by fitting the experimental (white squares) and numerical (black circles) vertical displacement fields with the equation (\ref{eq_1}) when varying the width of the intruder $D$.}\label{Fig3_Diffusion}
\end{figure}

We should emphasize that this model does not account for the size effects of the intruder since it assumed that the width of the intruder is very large compared to the diameter of the grains \cite{melo_kinematic_2008}. In the present situation, where successive blockages of the disks appear, we fit the displacement profiles varying the maximum vertical displacement at the intruder $\gamma_0$ and the diffusion parameter $D_p$. For both numerical simulations and experiments, the model describes well the long-term displacements of the disks [Figs. \ref{Fig2_LongTime}(c)-(d)], \AS{but for large $x$: as described latter, the formation of arches spreads the displacement fields laterally more than the prediction of the model}. The fitting parameter $\gamma_0$ increases with the width of the intruder, as shown in Fig. \ref{Fig3_Diffusion}(b), and we could expect a saturation for very large widths at the value of the withdrawal length $10\,d_g$.

The diffusion coefficient $D_p$ obtained by fitting the displacement profiles strongly increases with the width of the intruder and deviates from the values usually observed in silos (of few $d_g$) [Fig. \ref{Fig3_Diffusion}(a)]. However, the measurements performed on silos are very dispersed in the literature \cite{choi_velocity_2005}, but values of about $4 \, d_g $ are typically observed \cite{medina_velocity_1998}. Also, the model presented here is based on a continuous displacement field and is valid as long as the width of the orifice is small compared to the size of the system but large compared to the size of the disks. Here, we focus on the region close to the intruder and the grains are not very small compared to the intruder.

\section{Short-term rearrangements}  \label{Sec4}

\begin{figure}
\centering
  \includegraphics[width=0.48\textwidth]{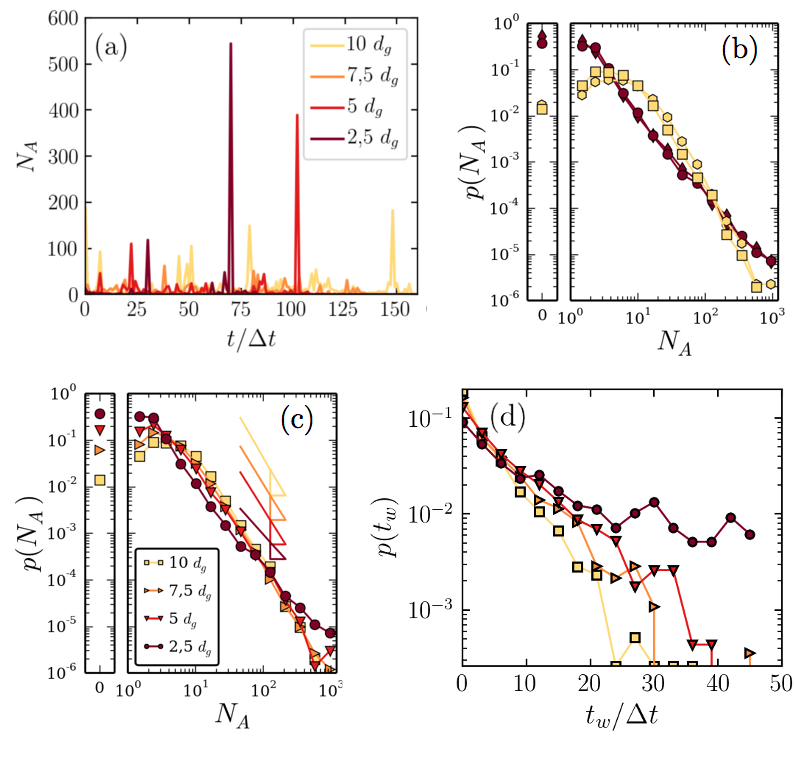}
 \caption{(a) Time sequences of avalanche sizes obtained numerically for different intruders widths (10, 7.5, 5 and 2.5 $d_g$). (b) Avalanche size distributions obtained numerically (circles and squares) and experimenlly (hexagons and diamonds) for two intruder widths (10 $d_g$: yellow, 2.5 $d_g$: brown). (c) Avalanche size distributions obtained numerically for four intruder widths. The slopes of the power laws are $-2.6$ (for $D=10\,d_g$), $-2.4$ (for $D=7.5\,d_g$), $-2.5$ (for $D=5\,d_g$), $-1.6$ (for $D=2.5\,d_g$). (d) Distributions of the waiting time between avalanches for which $ N_A > \langle N_A \rangle $. Color are the same than in (c).}\label{Fig4_ShortTerm}
\end{figure}

The main qualitative features observed with a large intruder \cite{merceron2016reorganization} are not significantly modified when reducing the intruder width and we still find a broad distribution of the rearrangement amplitude.  To characterize quantitatively the reorganization of the granular medium, we define the ava-lanche size $N_A$ as the number of disks whose absolute displacement is larger than the displacement of the intruder between two-time steps ($d_g/16$). We report in Fig. \ref{Fig4_ShortTerm}(a) the time-evolution of the avalanche size obtained numerically for different intruder widths. The intermittent dynamics is more pronounced when the intruder width, $D$, is small. Indeed, for the smallest intruder considered here, $D=2.5\,d_g$, the time-evolution of $N_A$ shows long period where no large rearrangement is observed punctuated by few events of very large amplitude, where $N_A$ is typically larger than 100. For the largest intruder, $D=10\,d_g$, the rearrangement events occur more frequently and have a smaller amplitude. This difference can be explained by stable configurations (arches) adopted by the granular packing for small intruder widths.
 As a result, the grains remain static during a longer period where the system accumulates potential energy since the intruder continues to move downward. When the arch becomes unstable, the energy is released in the form of an event impacting many grains and generating an avalanche of large size $N_A$. In the case of a large intruder, the probabilities to form a large arch are much smaller, and the occurrence of large events is smaller.

The difference observed for the rearrangement following the withdrawal of small or large intruders is also visible on the avalanche size distributions reported in Fig. \ref{Fig4_ShortTerm}(b), where the distributions obtained numerically and experimentally are in excellent agreement. We report in Fig. \ref{Fig4_ShortTerm}(c) the avalanche size distributions for four different intruder widths ($10\,d_g$ to $2.5\,d_g$). We always observe a power law distribution for the largest events, but the slope becomes less steep for the small intruders (from $-2.6$ to $-1.6$). \AS{The slope changes sharply when the intruders width becomes of the order of few grains diameters and suggest that an additional physical effect comes into play: the formation of arches}. Indeed, the probability of jamming is smaller for a large intruder, which results in a smaller probability of large avalanche size. It also leads to the apparition of a maximum of the distribution at a value close to the number of disks in contact with the tip of the intruder \cite{merceron2016reorganization}.  
Finally, the formation of arches modifies the global shape of the avalanche size distribution.
This observation, by analogy with the silos, seems to highlight the existence of a critical size beyond which jamming events become rare \cite{to_jamming_2001}.

The fundamental difference in the reorganization process for small enough intruder is visible in the distribution of waiting time between two avalanches of large size, \textit{i.e.}, for which $ N_A > \langle N_A \rangle $ (where $\langle N_A \rangle$ is the mean size of the avalanche in the system). The probability distribution function (PDF) of the waiting time $t_w$, where $t_w$ represents here the number of iterations when removing the mechanical intruder. \AS{A numerical iteration corresponds to one snapshot in the experiments. It is thus equivalent to non-dimensionalize the time in our system using a characteristic time $d_g/(16\,v)$, a relevant time scale for the trigger of the reorganizations}. Here, the PDF of the waiting time exhibits an exponential decrease for all width larger than $5 \, d_g$: the rearrangements remain decorrelated in time. For a smaller intruder width, the distribution of waiting times is closer to a power law (Fig. \ref{Fig4_ShortTerm}(d)). This observation suggests a correlation between events, induced by the dominant role of the formation of arches for $D=2.5\,d_g$. 


\section{Formation and breaking of arches}  \label{Sec5}

\begin{figure}
\centering
  \includegraphics[width=0.48\textwidth]{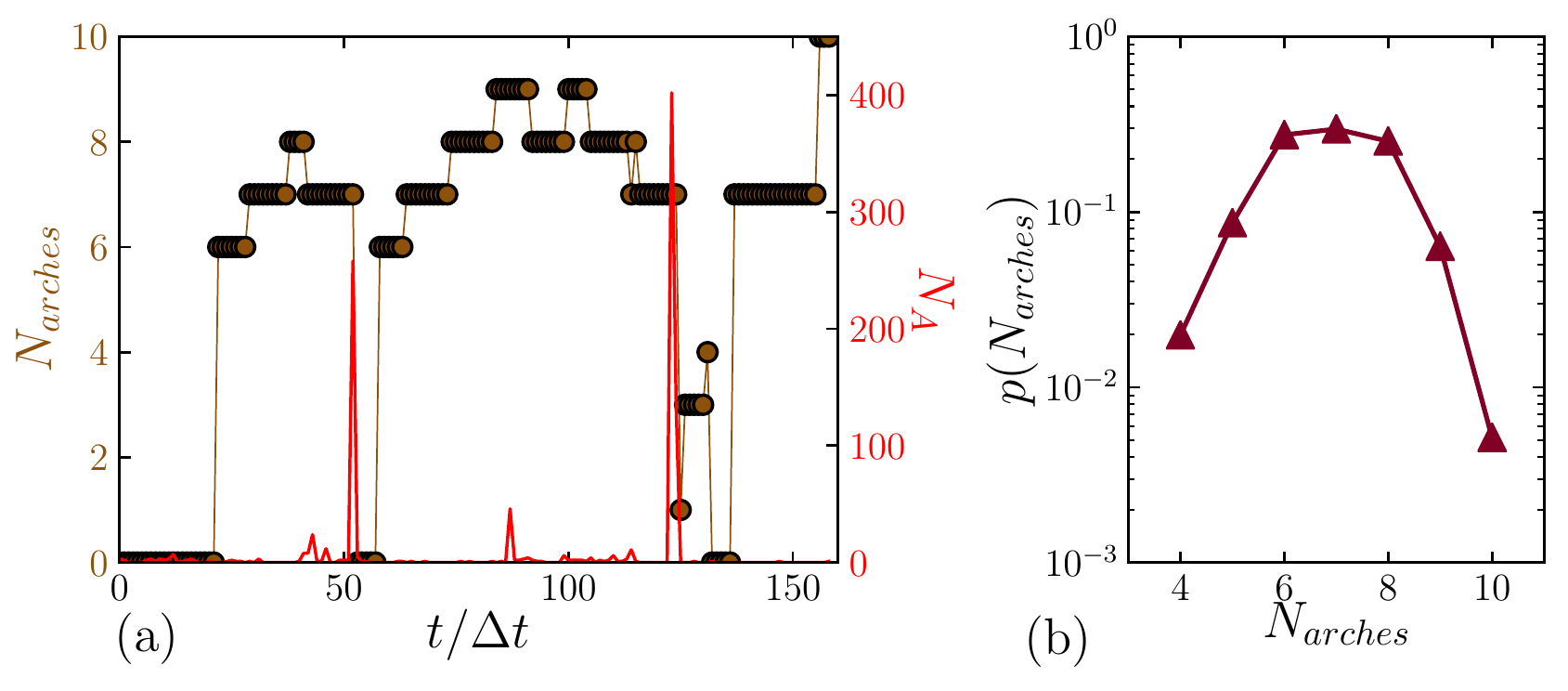}
	\includegraphics[width=0.26\textwidth]{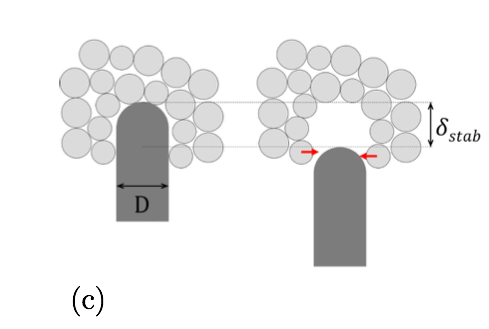}
	\includegraphics[width=0.21\textwidth]{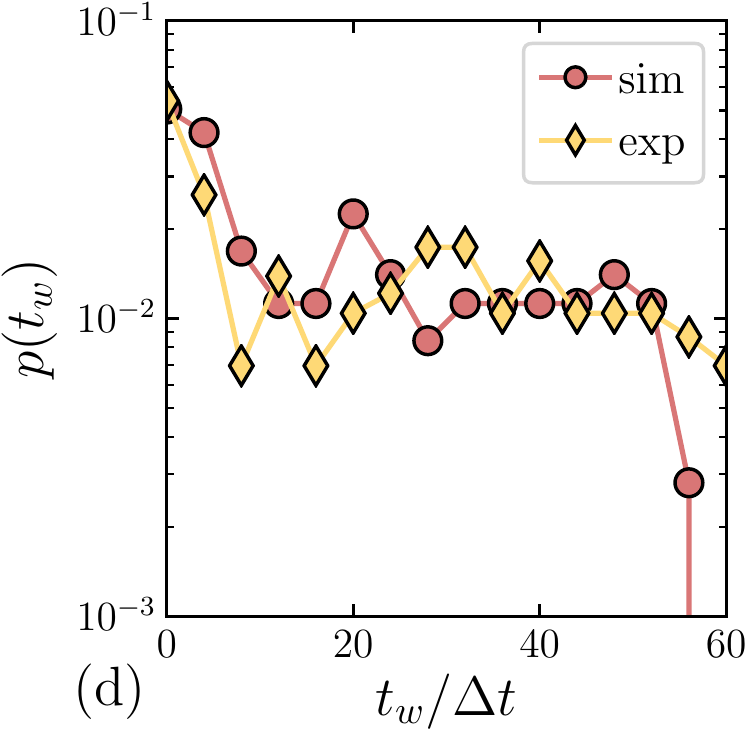}
 \caption{(a) Evolution of the number of disks in the arch $N_{arches}$ (brown circles) and avalanche size $N_A$ as a function of the dimensionless time for an intruder of width $2.5\,d_g$. (b) Distribution of the number of disks contained in the arches. (c) Schematic showing the possible formation and destabilization of an arch. (d) Time distributions between two events of sizes larger than $N_A=50$ (numerics: round, experiments: diamonds). }\label{Fig6_DistributionArches}
\end{figure}

The withdrawal of a small intruder of width $D=2.5\,d_g$ induces the formation of stable arches that block the flow of grains in the vicinity of the intruder. 
 While arches formed around outlets of silos can only be destabilizing by an external perturbation, withdrawing the intruder allows both the formation and later the destabilization of the arches.

The time-evolution of the number of disks forming the arch and the avalanche size $N_A$ is reported in Fig. \ref{Fig6_DistributionArches}(a). We observe that arches are formed after a delay corresponding to the first reorganizations around the intruder. Therefore, an arch follows here a process of formation, destabilization, and reformation. The destabilization of the mechanically stable arch quickly leads to the emergence of the next arch. The destabilization of the arches are related to the strong discontinuous dynamics observed for small intruder widths as shown in Fig. \ref{Fig6_DistributionArches}(a). Indeed, our results show that avalanches with large amplitudes ($N_A > 100$) are always associated with the complete or partial destabilization of an arch.

We report in Fig. \ref{Fig6_DistributionArches}(b) the probability distribution function (PDF) of the number of disks contained in the arches. This distribution has the appearance of an asymmetrical bell, similar to the arches forming at the outlet of a silo \cite{garcimartin_shape_2010}. Here, the number of disks in an arch varies from 4 to 11 disks with a large probability of size between $6$ and $8$ disks. We also observe that the centers of the disks forming the base of the arch are preferably separated by a distance between $2.5\, d_g$ and $3.5\, d_g$, which is consistent with the width of the intruder used here ($D=2.5\, d_g$). Most of the arches are formed on disks located on either side of the intruder\AS{, which is thus the main responsible for the geometry of the arches, their formation but also their destabilization.}
Because the arches are mostly generated by the intruder withdrawal, the destabilization of the arches likely is a periodic phenomenon induced by the withdrawal of the intruder at a constant velocity. The destabilization period would then be related to a displacement length of the intruder corresponding to the typical height of the arches [Fig. \ref{Fig6_DistributionArches}(c)].   \AS{The image on the left corresponds to the moment of formation of the arch, \textit{i.e.} when the disks start to be in contact with the tip of the intruder.}
 The arch remains stable while the intruder is moving downward, up to point where the disks at the base of the arch are overpassed by the tip of the intruder. The arch is then destabilized when the first disks of the base can move radially toward the intruder [Fig. \ref{Fig6_DistributionArches}(c)]. Considering the circular shape of the tip, we estimate the displacement of the intruder before destabilizing an arch to be between the radius and the diameter of the tip, $1.25\, d_g <\delta_{stab}< 2.5\, dg$, corresponding to $20<t_w/\Delta t< 40$ in the numerical simulation. 

We measure experimentally and numerically this period of the lifetimes of the arches in Fig. \ref{Fig6_DistributionArches}(d). Typically, the collapse of an arch corresponds to events with avalanche sizes greater than $50$. We, therefore, measured the waiting times between two reorganizations satisfying such condition which allows observing typically three arches during one experiment. 
 The distribution of the waiting time does not exhibit a strong periodicity but shows an increase in the occurrence of waiting times close to 30-40, which corroborates the mechanism explained above. Moreover, short waiting times are the most frequent and are the result of the collapse of an arch in several consecutive events.

\section{Conclusions}

In this paper, we have highlighted experimentally and numerically that the size of the intruder plays a crucial role in the rearrangements dynamics of a bi-dimensional granular media from which a vertical intruder is withdrawn. Decreasing the width of the intruder with respect to the size of the disks leads to a strong change in the nature of events: from moderate amplitudes and regular events, the reorganization becomes much more discontinuous in time. For the smallest intruders considered here, the energy brought by the local disturbance is stored in stable arches whose destabilization is self-sustained by the downward motion of the intruder and is accompanied by very strong events releasing the energy stored in the arches. 


%
%
    \bibliography{Biblio_EPJE_Merceron}
    \bibliographystyle{unsrt}

    \end{document}